\documentclass[onecolumn]{aastex631} 

\usepackage{natbib}
\usepackage{amsmath}
\usepackage{amssymb}
\usepackage{subfigure}
\usepackage{url}
\usepackage{CJK}
\usepackage{longtable}
\usepackage{graphicx}   
\usepackage{epsfig}
\usepackage{bm}
\usepackage{multirow}
\usepackage{color}  
\usepackage{booktabs}

\renewcommand{\vec}[1]{ {\mathbf #1} }

\newcommand{\pder}[2]{ \frac{\partial #1}{\partial #2} }
\newcommand{\grad}{ {\bf \nabla } }
\newcommand{\curl}{ {\bf \nabla} \times}

\newcommand{\Fig}{{Figure}}

\newcommand{\dive}{\nabla\cdot}

\shorttitle{Sun-to-Earth CME modelling}
\shortauthors{Jiang et al.}

\begin{document}

\title{Modeling of Coronal Mass Ejection Originated from a Sheared Arcade of Realistic Active-Region Scale and Its Propagation in the Heliosphere: Methodology}

\correspondingauthor{Chaowei Jiang} \email{chaowei@hit.edu.cn}

\author[0000-0002-7018-6862]{Chaowei Jiang}
\affiliation{State Key Laboratory of Solar Activity and Space Weather, School of Aerospace, Harbin Institute of Technology, Shenzhen 518055, China}

\author[0000-0001-8605-2159]{Xueshang Feng}
\affiliation{State Key Laboratory of Solar Activity and Space Weather, School of Aerospace, Harbin Institute of Technology, Shenzhen 518055, China}

\author{Liping Yang}
\affiliation{State Key Laboratory of Solar Activity and Space Weather, National Space Science Center, Chinese Academy of Sciences, Beijing 100190, China}

\author{Huichao Li}
\affiliation{State Key Laboratory of Solar Activity and Space Weather, School of Aerospace, Harbin Institute of Technology, Shenzhen 518055, China}

\author{Jinhan Guo}
\affiliation{School of Astronomy and Space Science and Key Laboratory of Modern Astronomy and Astrophysics, Nanjing University, Nanjing 210023, China}
\affiliation{Centre for mathematical Plasma Astrophysics, Department of Mathematics, KU Leuven, Celestijnenlaan 200B, B-3001 Leuven, Belgium}

\author{Pingbing Zuo}
\affiliation{State Key Laboratory of Solar Activity and Space Weather, School of Aerospace, Harbin Institute of Technology, Shenzhen 518055, China}

\author{Yi Wang}
\affiliation{State Key Laboratory of Solar Activity and Space Weather, School of Aerospace, Harbin Institute of Technology, Shenzhen 518055, China}

\begin{abstract}
  Simulating coronal mass ejections (CMEs) from their origin in active regions (ARs) to their propagation to Earth remains challenging, particularly when aiming to resolve AR scales and employ realistic magnetic field strengths without compromising computational efficiency. Here we present a methodology for end-to-end CME modeling that addresses these challenges. Three nested magnetohydrodynamic simulations are coupled to jointly cover the heliosphere from solar surface to beyond $1.5$~au. A block-structured adaptive mesh refinement scheme is employed to achieve $\sim 700$~km resolution in the low corona, allowing AR scales to be resolved while maintaining the total grid count below $10^8$ across the entire computational domain. A semi-relativistic Boris correction combined with a relativistic mass-density factor is used to handle magnetic field strengths up to $10^3$~G without prohibitively small time steps. Using this model, we simulate the emergence of a bipolar AR into the corona, the initiation of a CME by shearing of the AR core field and the subsequent evolution. Our simulation captures its pre-eruption energy buildup, triggering by magnetic reconnection, rapid acceleration, and propagation to 1~au and beyond. The simulated CME exhibits a three-part structure in synthetic coronagraph images and a torus-shaped flux rope in the heliosphere, with synthetic in-situ observations showing shock formation, density compression, and a prolonged southward $B_z$ component at 1~au. The entire simulation requires about one day on a moderately sized cluster (e.g., $600$ processors), while the simulated CME takes three days to arrive at 1~au, offering a lead time of two days if used for forecasting.
\end{abstract}


\section{Introduction}
\label{sect:intro}

Coronal mass ejections (CMEs) are among the most energetic eruptive phenomena in the solar atmosphere, characterized by the abrupt release of large-scale magnetized plasma into interplanetary space. Since their first detection by the coronagraph onboard NASA's Seventh Orbiting Solar Observatory (OSO-7) in 1971, CMEs have been a central subject in solar and heliospheric research due to their profound impacts on space weather and geomagnetic activity. Understanding the complete life cycle of CMEs from their initiation in the low corona to their propagation through the heliosphere and eventual interaction with Earth remains a fundamental challenge in solar and space physics~\citep{lugazImportanceFundamentalResearch2023,temmerCMEPropagationHeliosphere2023,torokLearnWalkYou2023}.

Over recent decades, substantial progress has been made through a combination of observational analyses, theoretical models, and increasingly sophisticated numerical simulations~\citep{alexanderBriefHistoryCME2006,forbesCMETheoryModels2006,linReviewCurrentSheets2015,manchesterPhysicalProcessesCME2017,luhmannICMEEvolutionInner2020,shenPropagationCharacteristicsCoronal2022}. Magnetohydrodynamic (MHD) simulations have proven indispensable for studying processes that are difficult to observe directly, such as magnetic reconnection in the low corona~\citep[e.g.,][]{wyperUniversalModelSolar2017,amariMagneticCageRope2018,dahlinModelEnergyBuildup2019,jiangFundamentalMechanismSolar2021,yanFastPlasmoidmediatedReconnection2022,guoThermodynamicMagneticTopology2023a}, flux-rope formation~\citep[e.g.,][]{aulanierFormationTorusUnstableFlux2010,guoSolarMagneticFlux2019,inoueFormationDynamicsSolar2018,heDatadrivenMHDSimulation2020,fanImprovedMagnetohydrodynamicSimulation2022,guoDatadrivenModelingCoronal2024,xingUnveilingInitiationRoute2024a}, and the interaction of CMEs with the structured solar wind~\citep[e.g.,][]{manchesterThreedimensionalMHDSimulation2004,lugazNUMERICALINVESTIGATIONCORONAL2011,lionelloMagnetohydrodynamicSimulationsInterplanetary2013,shenTurnSuperelasticCollision2016,jinDATACONSTRAINEDCORONALMASS2017a,torokSuntoEarthMHDSimulation2018,yangNumericalMHDSimulations2021,guoDependenceCoronalMass2024,linanCoronalMassEjection2024,caiMHDModellingNearsun2025,manchesterHighresolutionSimulationCoronal2025}. However, existing MHD simulations of CMEs often suffer from limitations: either they focus on the initiation phase within a local Cartesian domain of typical active region (AR) sizes, or they model the propagation phase with simplified initial conditions (often with an assumed eruptive flux rope such as the Gibson-Low model~\citep{gibsonTimeDependentThree1998}, the Titov-D{\'e}monlin model~\citep{1999A&A...351..707T} and its variant versions~\citep{titovRegularizedBiotSavart2018,titovMagnetogrammatchingBiotSavartLaw2024}, and the spheromak model~\citep{litesPossibleAscentClosed1995,singhApplicationModifiedSpheromak2020}) that do not self-consistently capture the magnetic-energy buildup prior to eruption. Some studies of CME generation and propagation~\citep[e.g.,][]{rileyUsingMHDSimulation2003,2003PhPl...10.1971L,talpeanuNumericalSimulationsShearinduced2020a} have considered the energy buildup by shearing a coronal helmet streamer and initiating the eruption with flux cancellation, but such source region of CME is far larger than typical ARs. Consequently, a seamless simulation that covers the entire journey of a CME from its birth near the solar surface of AR scale to its arrival at 1 au has remained elusive.

In our recent work \citep{caiMHDModellingNearsun2025}, we presented a global-corona MHD simulation of a CME originating from a sheared magnetic arcade, consistent with the fundamental mechanism of solar eruption initiations as demonstrated in \citet{jiangFundamentalMechanismSolar2021}. That study demonstrated that an internal current sheet forms gradually as driven by photospheric shearing motions, and fast reconnection at this sheet triggers the eruption, producing a classic three-part CME structure. Similar processes of CME initiation have also been shown in many earlier studies~\citep{mikicDynamicalEvolutionSolar1988,mikicDisruptionCoronalMagnetic1994,wolfsonShearinducedOpeningCoronal1995,choeEvolutionSolarMagnetic1996,antiochosModelSolarCoronal1999,amariCoronalMassEjection2003,rileyUsingMHDSimulation2003,talpeanuNumericalSimulationsShearinduced2020a}.
Nevertheless, our simulation~\citep{caiMHDModellingNearsun2025} was confined to the near-Sun region (up to $\sim 20R_{\odot}$), leaving the subsequent interplanetary evolution unexplored. Moreover, in that simulation, although an adaptive mesh refinement (AMR) grid is employed that dynamically adjusts the grid resolution based on evolving features \citep[{which is a common practice in CME simulations, e.g.,}][]{grothGlobalThreedimensionalMHD2000,lynchTopologicalEvolutionFast2008,karpenMechanismsOnsetExplosive2012,talpeanuNumericalSimulationsShearinduced2020,baratashviliEffectAdaptiveMesh2024,manchesterHighresolutionSimulationCoronal2025}, the finest grid is still relatively coarse ($0.25^\circ$ in Carrington latitude and longitude) which could not resolve the fine-scale magnetic structures within realistic ARs. Another limitation is that the magnetic field strength was artificially reduced as only a few tens of Gauss to avoid the severe time-step constraints imposed by high Alfv{\'e}n speeds, which is a common compromise in earlier global simulations that limits physical realism as typical ARs have magnetic field strength up to a few thousands Gauss. 

In the present study, we overcome these limitations by introducing two key technical advances to the model. First, we designed a set of AMR refinement criteria enabling a multi-scale simulation that captures both global solar-wind structures and fine-scale AR dynamics by a few of $10^7$ grid points in total. The highest resolution is concentrated in the AR, reaching $\sim 700$~km in the low corona comparable to the resolution of SDO/HMI magnetograms. This capability not only allows us to simulate ARs at reasonable physical scales for CME initiations  but also paves the way for future data-driven simulations that directly use high-resolution observational data. Second, we implement a semi-relativistic Boris correction combined with a relativistic inertial density factor depending on an artificial light speed, which effectively limits the effective Alfv{\'e}n speed and advoids naturally occurring plasma velocities which are too costly to include in strong-field, fast-reconnection regions. This approach allows us to use realistic active-region magnetic field strengths (up to several thousand Gauss) without incurring prohibitively small time steps. By using an appropriate artificial light speed, the simulation can preserve the physical acceleration and dynamics of eruptions while maintaining computational feasibility.

Building on these methodological advances, we simulate the entire Sun-to-Earth journey of a CME by coupling three nested MHD models: a high-resolution coronal model (CO) covering $1$--$11$ $R_{\odot}$, an interplanetary model (IP) from $9$ to $64$ $R_{\odot}$, and a heliospheric model (HE) from $50$ to $360$ $R_{\odot}$. This multi-scale approach maintains appropriate resolution and physical approximations in each domain while ensuring a seamless transition of the erupting structure across model boundaries, a concept also developed in the Space Weather Modeling Framework~\citep{tothSpaceWeatherModeling2005}. We introduce an AR through a boundary condition mimicking the flux-emergence process into a steady-state solar-wind background, then energize it with photospheric shearing motions. The resulting eruption is driven by magnetic reconnection in a current sheet that forms within the sheared arcade. The CME subsequently propagates through the coupled model system, allowing us to examine its near-Sun morphology, kinematic evolution, shock formation, and magnetic-structure transformation in the heliosphere.

This work represents a significant step toward a comprehensive, self-consistent numerical framework for simulating Sun-to-Earth CME events. By bridging the gap between corona-focused and heliosphere-focused models and incorporating advanced numerical techniques that enable both high resolution and physical realism, we aim to elucidate the physical processes that govern CME evolution over large spatial scales and to pave the way for future data-driven, space-weather forecasting models.

The remainder of this paper is organized as follows. Section~\ref{sec:method} describes the numerical model, including the modified MHD equations with the Boris correction, the grid configuration and AMR criteria, the boundary conditions, and the procedures for constructing the background solar wind along with emergence and energizing of AR. Section~\ref{sec:res} presents the simulation results, covering the pre-eruption energy buildup, the initiation of the CME, its near-Sun evolution, and its propagation through the inner and outer heliosphere to 1~au. The computational efficiency of the model is also assessed in this section. Finally, Section~\ref{sec:concl} summarizes the main conclusions and outlines directions for future work.

\section{Methods}
\label{sec:method}

\subsection{Modifications to the MHD equations}\label{sec:equ}

We first describe the non-dimensionalization of the MHD variables used in the numerical model. The normalization is based on typical values at the coronal base, as summarized in Table~\ref{table1}. Unless otherwise stated with physical units, all variables presented in the remainder of the paper are given in non-dimensionalized form.

\begin{table}
  \centering
  \begin{tabular}{lll}
    \hline \hline
    Variable &  Expression & Value \\
    \hline
    Density        & $\rho_{s}$ =  $n (m_p + m_e) $           
                   & $1.67 \times 10^{-16}$~g~cm$^{-3}$ \\
    Temperature    & $T_{s}$                                  
                   & $1\times 10^{6}$~K \\
    Length         & $L_{s} = R_{\odot } $                    
                   &  $6.96\times 10^2$~Mm\\
    Pressure       & $p_{s}=2nk_{B}T_{s} = \rho_{s}R T_{s}$   
                   &  $2.76\times 10^{-3}$~Pa \\
    Magnetic field & $B_{s}=\sqrt{\mu_{0}p_{s}}$              
                   & $5.89\times 10^{-1}$~G\\
    Velocity       & $v_{s}=\sqrt{p_s / \rho_{s} } = \sqrt{ R T_s }$ 
                   & $1.28 \times 10^2$~km~s$^{-1}$ \\
    Time           & $t_{s}=L_{s}/v_{s}$                      
                   & $5.42 \times 10^3$~s \\
    Energy         & $E_s = \rho_s v_s^2 L_s^3 $              
                   & $9.31\times 10^{30}$~erg \\
    \hline
  \end{tabular}
  \caption{Parameters used for non-dimensionalization. $n$ is a
    typical value of electron number density in the corona given by
    $n=1\times 10^{8}$~cm$^{-3}$. $m_p$ and $m_e$ are the proton and electron mass, respectively. $k_B$ is the Boltzmann constant. $R=2k_{B}/(m_p + m_e)$ is the gas constant.}
  \label{table1}
\end{table}

Simulating the global corona with an AR of realistic size and magnetic field strength presents a significant numerical challenge. On the one hand, ARs demand the highest resolution within the computational domain; in this study, we adopt a grid resolution comparable to that of HMI magnetograms, with a minimum cell size of $\Delta L_{\min} \approx 700$~km. On the other hand, the magnetic field in ARs can reach up to several $10^3$~G. In a typical coronal plasma with number density $n \approx 10^8$~cm$^{-3}$, the corresponding Alfv\'en speed is on the order of $10^5$~km~s$^{-1}$. For an explicit time-stepping scheme, this leads to a severely restricted time step:
\begin{equation}
\Delta t_{\min} \approx C \frac{\Delta L_{\min}}{v_{\rm A}} \approx 10^{-3}~{\rm s},
\end{equation}
assuming a Courant number of $C = 0.3$. Since CME evolution spans several days, such a time step is computationally prohibitive. Moreover, excessively small time steps can increase numerical dissipation and degrade solution accuracy.

To address this problem, we incorporate a modification to the MHD equations inspired by the Boris correction, a semi-relativistic approach commonly employed to alleviate the stringent time-step constraints imposed by large Alfv{\'e}n velocities \citep{boris1970physically,gombosiSemirelativisticMagnetohydrodynamicsPhysicsBased2002,matsumotoNewHLLDRiemann2019}. Specifically, we adopt the following set of modified MHD equations:
\begin{eqnarray}\label{eq:modelequ}
  \pder{\rho}{t} + \dive (\rho\vec v) = 0, 
  \nonumber \\
  \pder{(\rho_e\vec v)}{t} + 
  \dive (\rho_e\vec v\vec v + \overline{I}p_{\rm tot}-\vec B\vec B - \nu \rho_e \nabla\vec v) = 
  \vec f -\vec B\dive\vec B,
  \nonumber \\
  \pder{T}{t} + \dive (T\vec v) = (2-\gamma)T\dive\vec v,
  \nonumber \\
  \pder{\vec B}{t} + \curl (-\vec v \times \vec B ) + \grad(-\mu_{\rm d}\dive \vec B) = -\vec
    v\dive \vec B,
\end{eqnarray}
where $\rho$ is the plasma density, $\vec v$ the velocity, $\vec B$ the magnetic field, $T$ the temperature, and $p_{\rm tot} = p + B^2/2$ the total pressure. In the momentum equation, we replace the standard inertial density $\rho$ with an effective inertial density $\rho_e$ defined as
\begin{equation}
  \rho_e = \gamma_r (\rho + \rho_B)
\end{equation}
where $\gamma_r = 1/\sqrt{1-v^2/c^2}$ and $\rho_B = B^2/c^2$ with $c$ being an artificially adjustable light speed. The original Boris correction introduces the pseudo-density $\rho_B = B^2/c^2$ to moderate the Alfv{\'e}n speed. However, we find that in practice, particularly in reconnection outflow regions where the magnetic tension force dominates, the plasma velocity $v$ may still exceed the artificial light speed $c$, leading to numerical instability. To mitigate this, we further enhance the inertial density by incorporating the relativistic factor $\gamma_r$. This modification ensures both the effective Alfv{\'e}n speed and the plasma velocity being bounded by $c$ which is explained as follows. The effective Alfv{\'e}n speed, 
\begin{equation}
  v_{\rm A} = \frac{B}{\sqrt{\rho_e}} < \frac{B}{\sqrt{\rho_B}} = c
\end{equation}
remains bounded by $c$. In our code, the conserved quantity is the momentum density $\vec m = \rho_e \vec v$, from which the velocity is recovered. It follows that
\begin{equation}
  \gamma_r = \sqrt{1+\frac{m^2}{(\rho+\rho_B)^2c^2}}
\end{equation}
where $m = |\vec m|$ and
\begin{equation}
  \vec v = \frac{\vec m}{\rho_e} = \frac{\vec m c}{\sqrt{(\rho+\rho_B)^2c^2 + m^2}}.
\end{equation}
Thus, the plasma speed $v$ is guaranteed to stay below $c$ throughout the simulation.

The choice of the artificial light speed $c$ follows a simple principle: during the CME triggering and acceleration phase, $c$ should be set sufficiently large to avoid artificially suppressing the Lorentz-force-driven acceleration, while still being significantly higher than the CME speed to avoid artificially reducing the CME speed. Given that CME speeds are at most a few thousand km~s$^{-1}$ (corresponding to a normalized velocity of several tens), we set the maximum value of $c$ to $100$ in normalized unit. After the acceleration phase, CMEs gradually decelerate when interacting with the ambient solar wind, eventually approaching the background solar wind speed during its propagation. To enhance computational efficiency, $c$ is progressively reduced in accordance with this evolutionary trend, yet being always higher than the CME speed by at least three times. The specific values used at different stages will be described in the context of the simulation results.

A small kinetic viscosity $\nu$ is used for keeping numerical stability during the very dynamic phase of the simulated eruptions. Specifically, it is given as $\nu = 0.05\Delta L^2/\Delta t$ depending on the local spatial resolution $\Delta L$ and the local time step $\Delta t$ (our code used a variable timestep scheme depending on the refinement levels), and only restricted within the strong-magnetized regions with the original Alfv{\'e}n speed $v_{\rm A} = B/\sqrt{\rho} > 10$.

Our simulations are conducted in the heliocentric Carrington co-rotating frame, where the external force is given by $\vec f = \rho[ \vec g -\vec \Omega \times (\vec \Omega \times \vec r) - 2\vec \Omega\times \vec v]$. The angular velocity $\Omega$ corresponds to the sidereal Carrington rotation rate of the Sun, with a value of $2\pi/25.38$ radians per day.

For the thermodynamic description, we currently adopt a simple polytropic assumption, in which the energy (temperature) equation is closed by choosing a specific value for the ratio of specific heats $\gamma$. Different values of $\gamma$ are used in the three simulation domains, reflecting the distinct physical regimes of the corona and heliosphere. More realistic thermodynamic treatments will be incorporated in future model developments.

To control the numerical errors associated with the divergence of the magnetic field, source terms proportional to $\dive\vec B$ are included in both the momentum and induction equations following the approach of \citet{powellSolutionAdaptiveUpwindScheme1999}. Although these term are errors in themselves, they allow the $\dive\vec B$ to be transported with the flow speed and limit the errors if the $\dive\vec B$ flows off the grid.
In addition, a diffusion term of the form $\grad(-\mu_{\rm d}\dive \vec B)$ is introduced in the induction equation to further suppress divergence errors \citep{marder1987method}. The diffusion coefficient is set to $\mu_{\rm d} = 0.05\Delta L^2/\Delta t$.

\subsection{Grid settings and numerical solver}

While it is possible in principle to cover the entire domain from the solar surface to 1 au and beyond with a single set of grid, such an approach would be computationally inefficient due to the vast differences in spatial and temporal scales between the corona and the heliosphere. A common strategy to overcome this challenge is to divide the computational domain into multiple regions and couple them in a sequential manner, which has been widely adopted in global solar wind and CME modeling studies~\citep[e.g.,][]{rileyModelingInterplanetaryCoronal2006,fengValidation3DAMR2012,jinDATACONSTRAINEDCORONALMASS2017a,torokSuntoEarthMHDSimulation2018,linanCoronalMassEjection2024}. Following this approach, we divide the computational domain into three distinct but coupled models along the radial direction. The first is the coronal model (hereafter CO), which spans from the solar surface to approximately $11R_{\odot}$, a region where the solar wind flow is predominantly supersonic and super-Alfv{\'e}nic. This radial range is smaller than that used in many other global coronal models (which often extend to $20R_{\odot}$ or beyond), and is chosen deliberately to reduce the computational cost of the CO model, which is the most time-consuming component of the coupled system due to the presence of strong magnetic fields and the need for high resolution near the solar surface. The second model, referred to as the interplanetary model (IP), extends from $9R_{\odot}$ to $64R_{\odot}$, a range also chosen for computational efficiency. It uses time-dependent MHD data extracted from the CO model at $9R_{\odot}$ as its inner boundary condition. The third model, termed the heliosphere model (HE), covers the region from $50R_{\odot}$ to $360R_{\odot}$ and is driven by data extracted from the IP model at $50R_{\odot}$. Each model employs a different value of the polytropic index $\gamma$ to reflect the dominant physical processes in its respective domain. Specifically, we set $\gamma = 1.05$ in the CO model to approximate a nearly isothermal corona, $\gamma = 1.3$ in the IP model, and $\gamma = 1.5$ in the HE model to account for the increasingly adiabatic behavior of the solar wind in the outer heliosphere.

For the angular coordinates, we adopt a so-called Yin-Yang grid to avoid the pole singularities inherent in standard spherical grids that cover the full sphere with $\theta \in [0,\pi]$ and $\phi \in [0,2\pi]$. The Yin-Yang grid consists of two identical, low-latitude partial spherical grids oriented perpendicular to each other. These two components overlap slightly along their boundaries to provide complete coverage of the sphere without singularities. A detailed description of the grid configuration can be found in \citet{jiangNEWCODENONLINEAR2012}.

The simulation employs a block-structured AMR technique, which dynamically adjusts the grid resolution in response to evolving physical features, thereby improving both accuracy and computational efficiency. All three models are implemented on AMR grids, each with different maximum refinement levels.

For the CO model, we use six levels of refinement. The base resolution is set to $\Delta\theta = \Delta\phi = 2^{\circ}$ in longitude and latitude, and the highest refinement level reaches $2^{\circ}/2^5 = 0.0625^{\circ}$, which corresponds to a spatial resolution of approximately $700$~km in the low corona. In the radial direction, the grid spacing is chosen such that $\Delta r = r\Delta\theta$, ensuring that the cells remain approximately cubic throughout the domain. At each time step, the mesh is refined or coarsened based on a set of criteria designed to capture regions of interest with sufficient resolution. These criteria are as follows.

The first criterion is based on a decomposition of the Lorentz force into the magnetic pressure gradient and magnetic tension force. We define a refinement indicator as
\begin{equation}
T_{\rm Lorentz} = \frac{(|\vec P_B| + |\vec T_B|)\sqrt[3]{\Delta V}}{(\rho+\rho_B)C_{\rm Lorentz}},
\end{equation}
where $\vec P_B = \grad (B^2/2)$ represents the magnetic pressure gradient, $\vec T_B = \vec B \cdot \grad \vec B$ the magnetic tension force, $\Delta V$ the cell volume, and $C_{\rm Lorentz}$ an adjustable parameter set to $200$ in this study. A grid block (comprising $6^3$ cells) is marked for refinement if any cell within it satisfies $T_{\rm Lorentz} > 1$, and refinement continues until the maximum allowed level is reached. This criterion is designed to ensure that regions with strong magnetic fields, large field gradients, or intense current densities--characteristic of eruptive ARs--are resolved with the finest grid resolution available.

The second set of refinement criteria is based on gradients in the mass density and on the divergence and curl of the momentum density $\rho_e \vec v$. We define two corresponding indicators:
\begin{equation}
T_{\rho} = \frac{|\grad\rho|\sqrt[3]{\Delta V}}{\rho C_{\rho}}, \qquad
T_{\rm M} = \frac{(|\dive (\rho_e \vec v)| + |\curl (\rho_e \vec v)|)\sqrt[3]{\Delta V}}{(\rho v_s + |\rho_e \vec v|)C_{\rm M}},
\end{equation}
where $v_s = \sqrt{p/\rho}$ is the sound speed, and $C_{\rho}$ and $C_{\rm M}$ are free parameters both set to $0.3$ in this study. A grid block is flagged for refinement if any cell within it satisfies $T_{\rho} > 1$ or $T_{\rm M} > 1$. These criteria are intended to capture dynamic features associated with the eruption and propagation of the CME. To maintain computational affordability, refinement based on these criteria is capped at a maximum resolution of $0.5^{\circ}$, as CME-driven structures such as fast ejecta and shocks often expand to scales comparable to the entire CO domain. 

Conversely, a grid block is considered for coarsening only when all three refinement indicators in all its cells fall below $0.2$.

For the IP and HE models, we employ four levels of refinement, with a base resolution of $2^\circ$ and a maximum resolution of $0.25^\circ$. The same refinement and coarsening criteria described above are applied to these models. Specially, in blocks adjacent to the inner boundary, the resolution is fixed at $0.5^\circ$ to ensure consistency with the boundary driving data, which are extracted from the inner model at a uniform resolution of $0.5^\circ$.

The MHD equations are solved using the space-time conservation element and solution element (CESE) method developed by \citet{jiangAMRSimulationsMagnetohydrodynamic2010b}, which has been extended to general curvilinear coordinates and implemented on block-structured AMR grids with the PARAMESH toolkit~\citep{macneicePARAMESHParallelAdaptive2000,olsonOverviewPARAMESHAMR2005}. The PARAMESH package has been further modified to accommodate multi-block overlapping grid systems such as the Yin-Yang grid~\citep{jiangNEWCODENONLINEAR2012,zhangParallelOverlappingCurvilinear2013}. The method treats space and time as a unified entity, enforces local flux conservation without Riemann solvers, and introduces spatial derivatives of conserved variables as independent unknowns for sub-cell resolution. An improved version of the scheme with predictor-corrector time integration is adopted to facilitate coupling with PARAMESH and simplify AMR operations. The solver has been validated through various MHD benchmark problems in both Cartesian and curvilinear coordinates.

\subsection{Boundary conditions}

At the inner boundary of the CO model, which corresponds to the solar surface, the plasma density and temperature are held fixed. The magnetic field at the boundary is updated by solving the induction equation, under the assumption that the electric field $\vec E_s$ is prescribed. When a surface driving velocity $\vec v_d$ is specified, the corresponding electric field is given by $\vec E_s = -\vec v_d \times \vec B_s$. In the absence of driving, we set $\vec v_d = \mathbf{0}$, which yields $\vec E_s = \mathbf{0}$. Alternatively, in cases where the electric field $\vec E_s$ is provided directly, the surface velocity can be recovered via $\vec v_s = (\vec E_s \times \vec B_s)/B_s^2$. Here, the subscript $s$ denotes quantities evaluated at the solar surface.

To specify the inner boundary conditions for the IP model, we extract from the CO model the full set of MHD variables, namely $\rho^{\rm CO}(\tau_n)$, $\vec v^{\rm CO}(\tau_n)$, $p^{\rm CO}(\tau_n)$, and $\vec B^{\rm CO}(\tau_n)$, on the spherical surface at $9R_{\odot}$. The data are sampled with a uniform spatial resolution of $0.5^{\circ}$ in both latitude and longitude at each CO time step $\tau_n$, which results in a non-uniform temporal cadence. At each IP time step $t_m$, we perform a time interpolation of these variables with third-order accuracy to obtain $\rho(t_m)$, $\vec v(t_m)$, $p(t_m)$, and $\vec B(t_m)$.

For the plasma variables ($\rho$, $\vec v$, $p$) and the radial magnetic field component $B_r$, the interpolated values are directly imposed at the inner boundary of the IP model. To update the tangential magnetic field components $B_{\theta}$ and $B_{\phi}$, we first compute the electric field as $\vec E = -\vec v \times \vec B$, and then evolve the field using the induction equation applied at the boundary. If the IP simulation time $t_m$ exceeds the time range covered by the CO output (for instance, after the CO model has reached a steady state), the boundary values for $\rho$, $\vec v$, $p$, and $B_r$ are held fixed, and the electric field is set to zero.

The inner boundary conditions for the HE model are prescribed in an analogous manner. MHD variables are extracted from the IP model on the spherical surface at $50R_{\odot}$ with a uniform resolution of $0.5^{\circ}$, and are used to drive the HE simulation following the same procedure described above.

For the outer boundary of all three models, the plasma variables $\rho$, $\vec v$, and $p$ are extrapolated radially using a zero-gradient condition. The magnetic field $\vec B$ is updated by solving the induction equation directly on the boundary surface.

\begin{figure*}
	\centering
  \includegraphics[width=\textwidth]{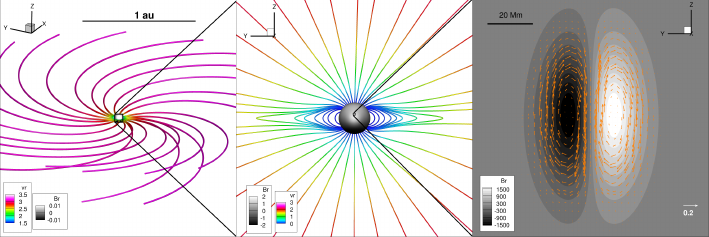}
	\caption{The initial steady state of the background solar wind with the embedded AR. Left panel: Magnetic field lines in the heliosphere extending to approximately $1.5$~au, colorcoded by radial velocity. The central sphere (indicated by the boxed region) represents the surface at $10R_\odot$ and is color-coded by the radial magnetic field $B_r$. The line segment and the accompanying number indicate the spatial scale. Middle panel: Magnetic field lines in the CO model, extending from the solar surface (color-coded by $B_r$) to $10R_\odot$. The small box outlines the location of the AR. Right panel: Enlarged view of the AR magnetic flux distribution. The arrows indicate the driving flow pattern used to inject free magnetic energy into the coronal field. All quantities are given in normalized units unless otherwise stated. The line segments with numbers in the panels show the spatial scales.}
	\label{fig:ARinBG}
\end{figure*}

\subsection{Construction of a background steady-state solar wind}
We first construct a steady-state solar wind background to initialize all three models. The plasma is initially specified using Parker's classical spherically symmetric solar wind solution, assuming an ideal adiabatic gas with $\gamma = 1.05$. At the solar surface, the density and temperature are set to $\rho = 0.33$ and $T = 1.8$ in normalized units. These values are chosen such that the simulated plasma number density and flow speed at 1~au are approximately $n \approx 10$~cm$^{-3}$ and $v \approx 400$~km~s$^{-1}$, which are typical of the slow solar wind observed at Earth. The magnetic field is initialized as a potential field and consists of a background dipole that represents the global coronal magnetic structure during solar minimum, with a maximum field strength of $1.5$~G at the poles.~\footnote{We note that this polar field is much lower low than typical strength (around $5 \sim 10$~G) in observations, and more realistic values will be considered in future studies.}

The CO model is integrated in time until a relaxed steady state is achieved. The time series of MHD variables extracted at the $9R_{\odot}$ sphere is then used to drive the IP model. Once the IP model reaches a steady state, the data at $50R_{\odot}$ are in turn used to drive the HE model to a steady state. During this background relaxation phase, the artificial light speed is set to $c = 100$ in the CO model and $c = 30$ in both the IP and HE models. The resulting steady-state solution is illustrated in the left two panels of \Fig~\ref{fig:ARinBG}.

\subsection{Emergence of AR}
We introduce an AR to the background mimicking a flux emergence process using an electric-field driving approach. Let $B_r^a(\theta,\phi)$ denote the surface flux density of the AR, and $B_r^g$ for the background flux. The flux emergence process is designed such that the surface radial field evolves in $t \in [0, t_0]$ by 
\begin{equation}
  B_r(t) = B_r^g + B_r^a \frac{e^{k t}-1}{e^{k t_0}-1},
\end{equation}
where $k$ is a constant controlling the flux emerging rate. To realize the flux emergence, we derive an inductive electric field $\vec E_I$ that is required to specify at the inner boundary. It has the form of 
\begin{equation}
  \vec E_I = \nabla_h \times \Phi \vec e_r = \left(\frac{1}{\sin\theta}\pder{\Phi}{\phi}, -\pder{\Phi}{\theta}\right).
\end{equation}
Applying the induction equation
\begin{equation}
  \pder{B_r}{t} = B_r^a\frac{ke^{k t}}{e^{k t_0}-1} =  -\nabla_h \times \vec E_I = \nabla_h^2 \Phi ,
\end{equation} 
which gives
\begin{equation}
  \vec E_I (t) = \vec E_{I}^a \frac{k e^{k t}}{e^{k t_0}-1},
\end{equation}
where $\vec E_{I}^a$ is recovered by solving $\nabla_h \times \vec E_I^a = -B_r^a$ (i.e., first solving $\nabla_h^2 \Phi^a = B_r^a$ and then $\vec E_I^a = \nabla_h \times \Phi^a \vec e_r$). We used $t_0 = 0.5$, $k=9.2$ and thus $e^{kt_0} \approx 100$. In this way, the surface velocity corresponding to the flux emergence, $\vec v = \vec E_I(t)/B_r(t)$, is roughly constant. In this phase, the artificial light speed is given as $c=100$.

The surface flux distribution of the AR is given by the sum of two 2D Gaussian functions on the $(\theta,\phi)$ plane,
\begin{equation}
	B_r^a = B_0 \exp\left(-\frac{\xi^2}{\sigma_\xi^2}\right)
  \left[ 
    \exp\left( -\frac{(\eta - \eta_0)^2}{\sigma_\eta^2}\right) - 
    \exp\left( -\frac{(\eta + \eta_0)^2}{\sigma_\eta^2}\right)
  \right],
\end{equation}
where $\xi = \theta - 5\pi/6$, $\eta=\phi-\pi$, $\eta_0 = \pi/1500$, $\sigma_\eta = \pi/150$, $\sigma_\xi = 2 \sigma_\eta$. The scale factor $B_0$ is given such that the maximal value of $B_r^a$ is 1000~G. As a result, the total unsigned magnetic flux of the AR is $1.7\times 10^{22}$~Mx. Therefore it has size and field strength of typical ARs. The magnetic flux distribution is shown in \Fig~\ref{fig:ARinBG}. The Carrington longitude and latitude of AR's center are $180^\circ$ and $30^\circ$, respectively.

\subsection{Surface driving}
After the AR is emerged, the simulation time is reset to $t=0$, and beginning with this state as the initial condition, we apply a surface rotation flow at each polarity of the AR to inject free magnetic energy into the AR. The driving flow is incompressible with streamlines aligning with the contour lines of $B_r$, therefore not altering the profile of $B_r$ on the surface. Specifically, it is given by
\begin{equation}
  \label{velocity}
  v_{r}=0,
    v_\theta = \frac{1}{R \sin \theta} \frac{\partial \Psi(B_r)}{\partial \phi},
    v_\phi = -\frac{1}{R} \frac{\partial \Psi(B_r)}{\partial \theta}
\end{equation}
with $\Psi$ given by
\begin{equation}
    \Psi = k B_r^2 e^{-(B_r^2 - B_{r, \mathrm{max}}^2)/B_{r, \mathrm{max}}^2}
\end{equation}
where $B_{r, \mathrm{max}}$ is the maximum value of $B_r$ at the surface, and $k$ is a scaling constant chosen so that the maximum surface velocity is $0.4$. The pattern of the flow is shown in the rightmost panel of \Fig~\ref{fig:ARinBG}. During this phase the artificial light speed is fixed as $c=100$. Therefore, the surface driving speed, although being much larger than the realistic photospheric values, is smaller than the effective Alfv{\'e}n speed (which approaches the artificial light speed $c=100$ when $B>100$) in the model by three orders of magnitude, and still represents a quasi-static driving to the model. A benefit of adopting such a relatively high driving speed is the significant reduction in computational time required to reach the eruption onset. In reality, the slow accumulation of magnetic free energy in the corona occurs over timescales of hours to days, driven by photospheric flows of only a few km~s$^{-1}$. By enhancing the driving speed while preserving the quasi-static nature of the evolution, we are able to compress the energy buildup phase substantially without altering the essential physical processes leading to the eruption. This acceleration of the pre-eruption evolution is essential for making global, high-resolution simulations of CME initiation computationally feasible.

\begin{figure*}
	\centering
  \includegraphics[width=0.8\textwidth]{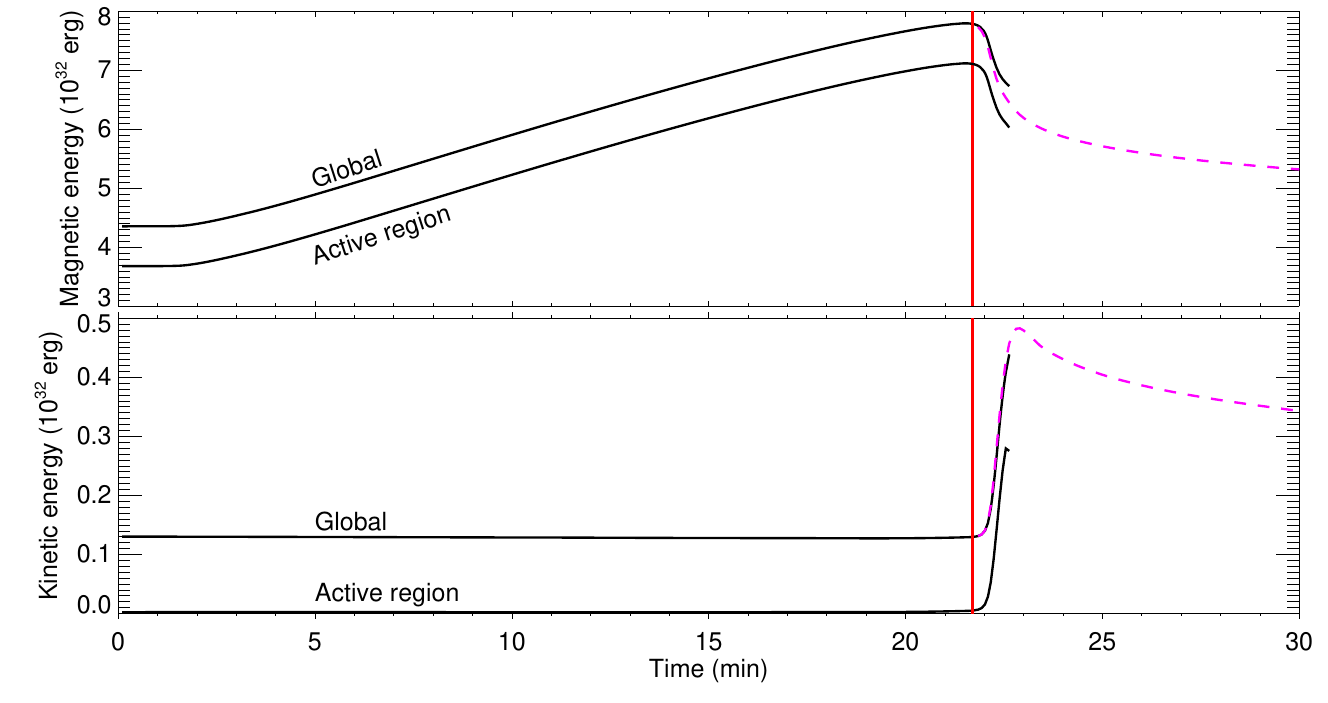}
	\caption{Evolution of magnetic and kinetic energies in the CO model. The global values are integration in the whole computational volume. The values of AR is integrated for a sub-volume with $30^\circ$ in both longitude and latitude centered at the AR and $r \in [1, 1.5]R_\odot$ in radial direction. The red vertical lines denote the onset time of the eruption. The dashed curves are global values starting at the eruption onset time with the bottom surface driving motion switched off.}
	\label{fig:erg}
\end{figure*}

\begin{figure*}
	\centering
  \includegraphics[width=\textwidth]{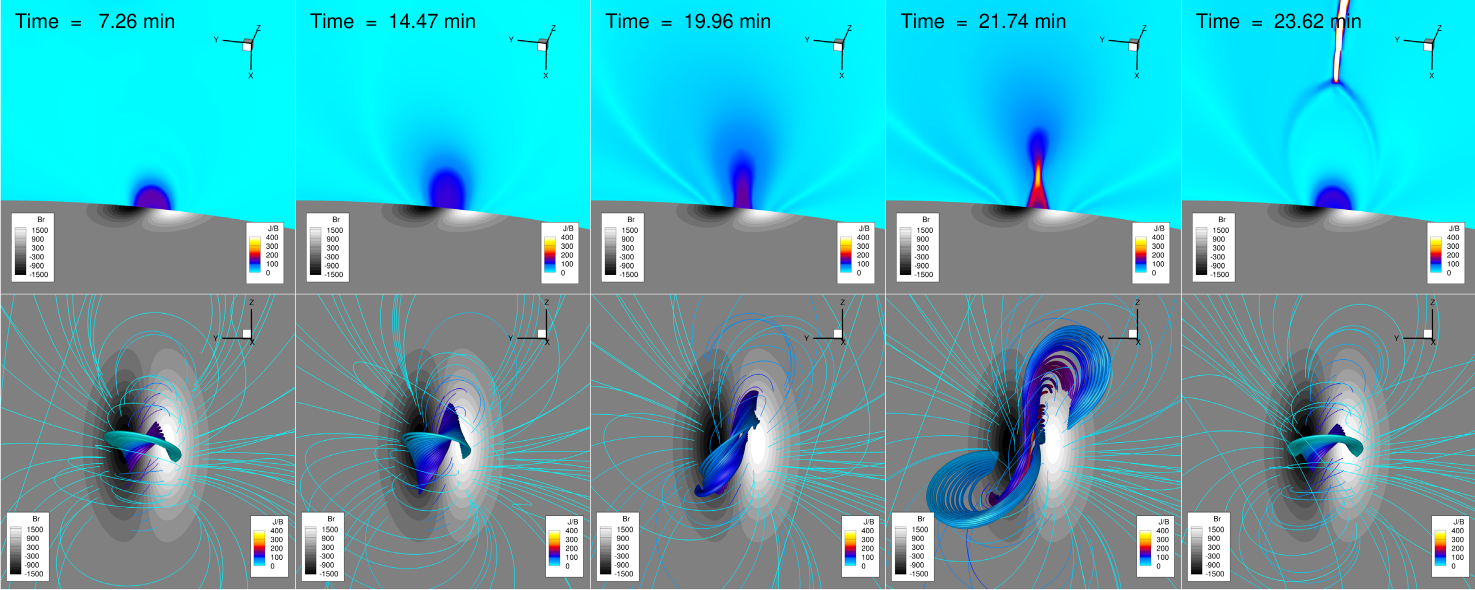}
	\caption{Evolution of the AR driven by the imposed surface flow. Top panel: Central cross section of the normalized current density $J/B$, with the solar surface at the bottom color-coded by the radial magnetic field $B_r$. Bottom panel: Magnetic field lines color-coded by $J/B$, where thick lines represent the core field and thin lines denote the enveloping field. The solar surface is shown in the background, color-coded by $B_r$. From left to right, the first three columns correspond to times before the eruption onset, while the last column shows the state shortly after the eruption begins.}
	\label{fig:preerup}
\end{figure*}

\section{Results}\label{sec:res}

\subsection{Pre-eruption evolution to CME initiation}

The evolution of magnetic and kinetic energies in the CO model, shown in \Fig~\ref{fig:erg}, clearly indicates that the eruption is initiated at approximately $t = 22$~min. Prior to eruption, the kinetic energy of the active region remains very low, confirming that its evolution is quasi-static. The initial magnetic energy of the active region is $3.7\times 10^{32}$~erg. Through the imposed surface driving, the total magnetic energy increases to about $7\times 10^{32}$~erg, implying a free magnetic energy of roughly $3.3\times 10^{32}$~erg, which is comparable to the energy of the initial potential field.

Following the onset, the magnetic energy decreases rapidly while the kinetic energy rises sharply. To assess the role of continued driving after eruption onset, we performed an additional simulation in which the surface driving was turned off once the eruption began. The resulting energy evolution, shown by the dashed curves in \Fig~\ref{fig:erg}, demonstrates that the eruption proceeds independently of the driving once initiated. The kinetic energy reaches its peak within a few minutes after the onset. During this acceleration phase, the magnetic energy decreases by approximately $2\times 10^{32}$~erg, of which about $20\%$ (or $0.37\times 10^{32}$~erg) is converted into kinetic energy. Thereafter, the kinetic energy gradually declines as the CME decelerates through interaction with the ambient solar wind.

Similar to our previous studies \citep{jiangFundamentalMechanismSolar2021,bianNumericalSimulationFundamental2022,bianHomologousCoronalMass2022}, the pre-eruption evolution of the AR is characterized by the gradual formation of a current sheet within the increasingly sheared magnetic arcade, driven by the imposed surface motion. This development is illustrated in \Fig~\ref{fig:preerup}, which shows the evolution of current density and magnetic field lines. Once the current sheet becomes sufficiently thin, tether-cutting reconnection \citep{mooreOnsetMagneticExplosion2001} initiates, leading to the subsequent eruption and the CME formation.


\begin{figure*}
	\centering
  \includegraphics[width=\textwidth]{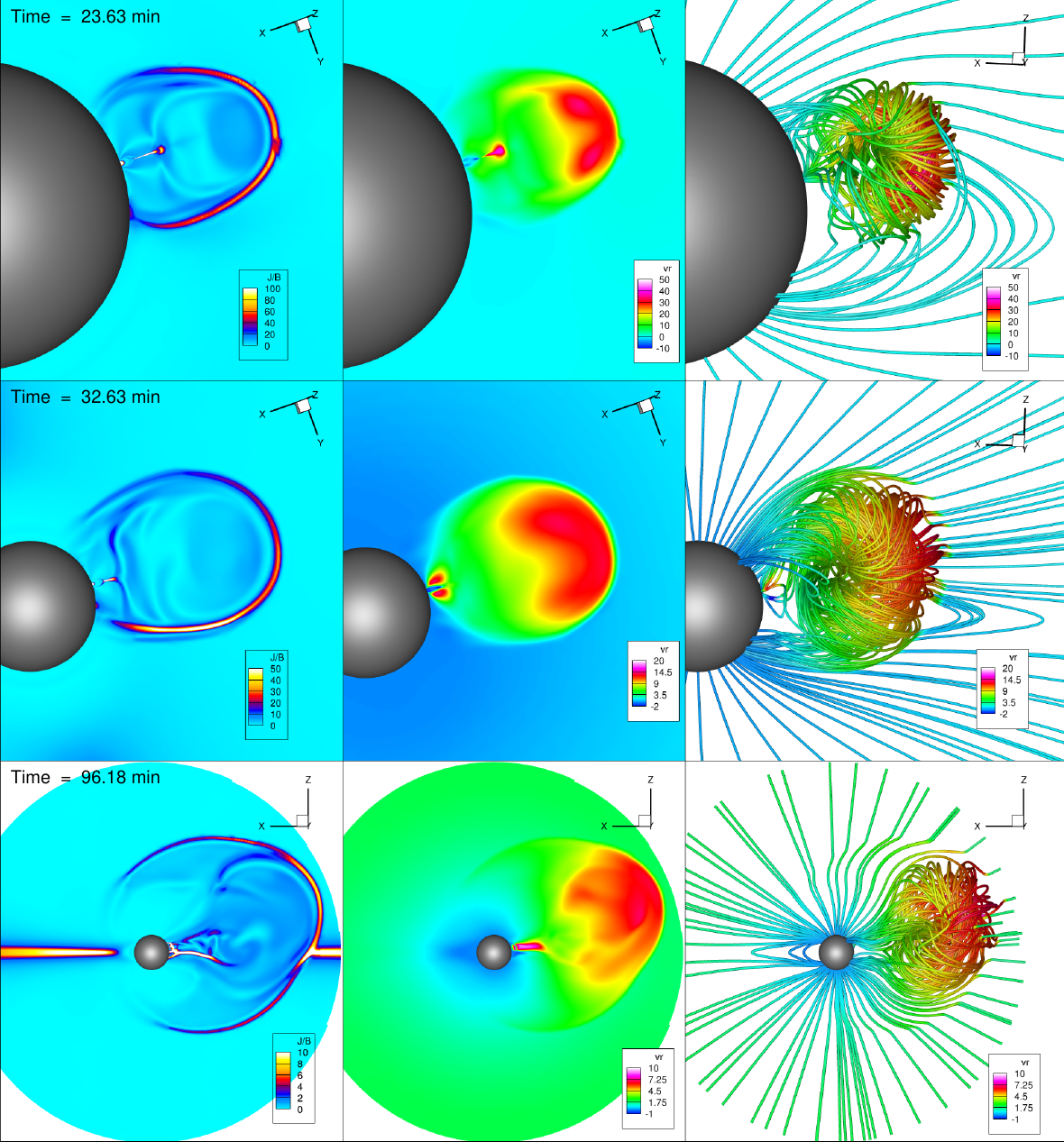}
	\caption{Evolution of the erupting magnetic structure near the Sun. From left to right, the panels show the normalized current density $J/B$ and the radial velocity $v_r$ on a central cross section through the AR, as well as the magnetic field lines at selected times during the eruption. The times are denoted in the left column from top to bottom. The color bars indicate the magnitude of $J/B$ and $v_r$ in normalized units. The formation of a twisted flux rope, the underlying current sheet, and the fast reconnection outflow are clearly visible. An animation of this evolution is available online, which shows the whole duration of the CO simulation. In the animation, the panels from left to right are magnetic field lines, the logarithm of plasma density $\log\rho$, and $J/B$. The top row is shown for the $y=0$ slice and the bottom row for $z=0$ slice.}
	\label{fig:AR_erup}
\end{figure*}

\begin{figure*}
	\centering
  \includegraphics[width=\textwidth]{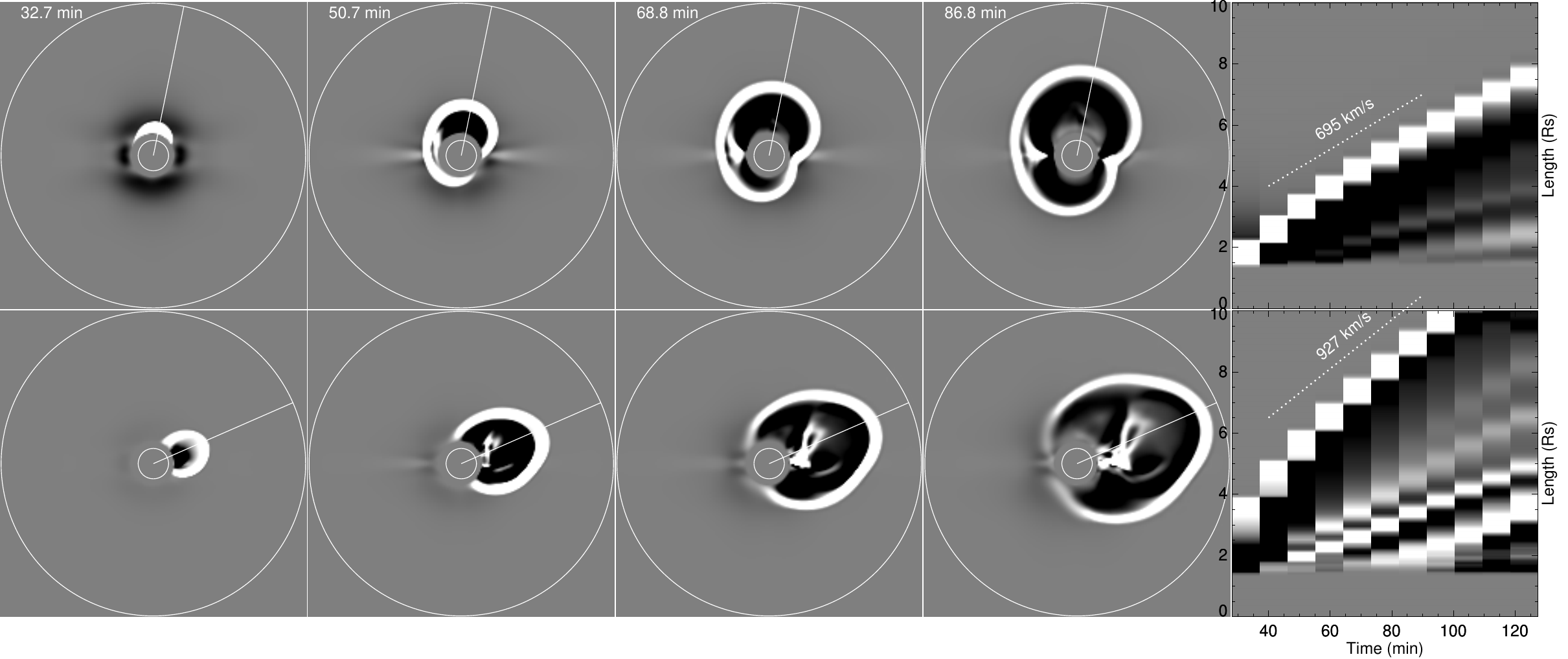}
	\caption{Running difference of synthetic coronal polarized brightness at two different view angles. Top row: face-on view, where the inner circle denotes the solar surface and the outer circle indicates $10R_\odot$. The rightmost panel shows a time stack along the direction indicated in the left panels, with the speed of the CME leading edge labeled. Bottom row: limb view.}
	\label{fig:pB}
\end{figure*}

\subsection{Near-Sun evolution of CME}

\Fig~\ref{fig:AR_erup} and the accompanying animation illustrate the near-Sun evolution of the erupting structure. The figure presents a central cross section of the normalized current density $J/B$ and the radial velocity $v_r$, overlaid with three-dimensional magnetic field lines. The density distribution on the same cross section is also shown in the animation.

During the eruption, a highly twisted and rapidly expanding magnetic flux rope forms as the main body of the CME. The leading edge of the eruption accelerates quickly, reaching speeds over $3000$~km~s$^{-1}$ shortly after onset, before decreasing to around $1500$~km~s$^{-1}$ by $t = 10$~min. Because the CME propagates much faster than the ambient solar wind, a fast shock develops ahead of the flux rope. Beneath the rope, an elongated current sheet extends downward, within which continuous magnetic reconnection drives high-speed reconnection outflows.

\Fig~\ref{fig:pB} displays running difference images of the synthetic coronal polarized brightness (pB) from two viewing angles, mimicking coronagraph observations. In the face-on view (top row), the eruption appears as a halo CME, characterized by a bright leading edge surrounding a dark cavity. At $10R_s$, the speed of the CME front remains around $700$~km~s$^{-1}$. In the limb view (bottom row), the CME exhibits the classic three-part structure: a bright front, a dark cavity, and a bright core. The leading edge speed in this view reaches approximately $930$~km~s$^{-1}$ at $10R_s$. This comparison shows that when viewed from different angles, the apparent speed of the CME also differs significantly.

The disturbance caused by the CME extends throughout the corona, indicating the global nature of the eruption. The leading edge of the CME takes on a roughly ellipsoidal shape that encloses the Sun, consistent with the expanding frontal structure observed in large-scale eruptions.



\begin{figure*}
	\centering
  \includegraphics[width=0.8\textwidth]{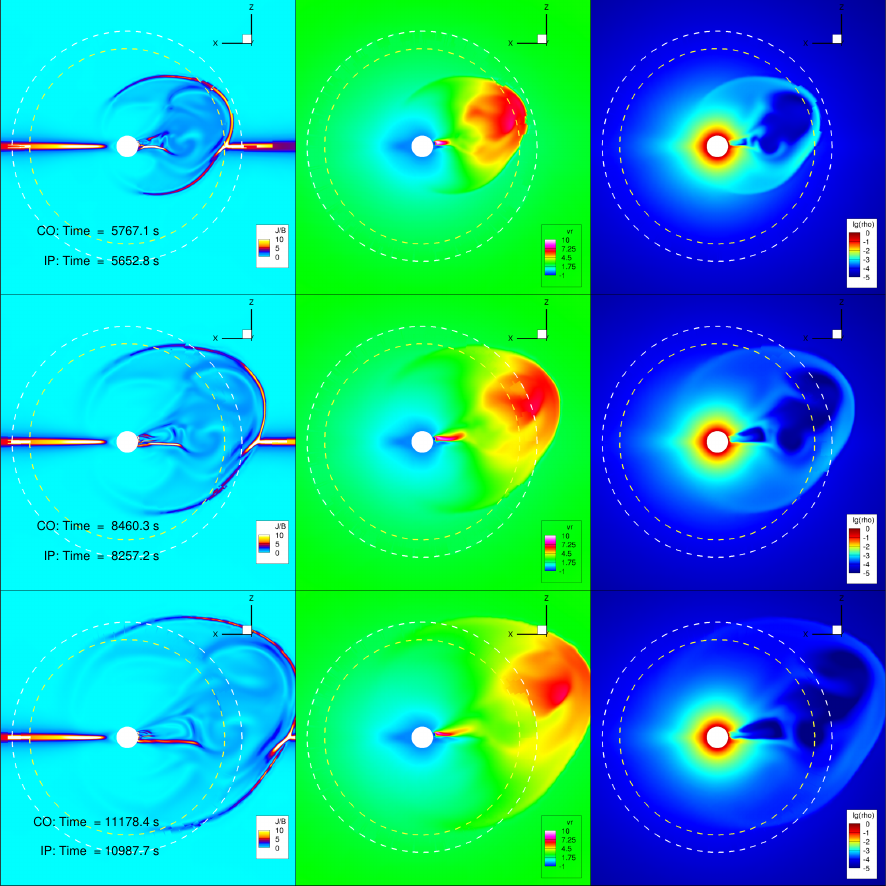}
	\caption{CME propagation through the interface between the CO and IP models at three different times (from top to bottom rows). From left to right, the panels show the normalized current density $J/B$, the radial velocity $v_r$, and the logarithm of density $\log(\rho)$ on the $y=0$ meridional plane. The innermost hollow circle represents the Sun. The dashed yellow circle at $r=9$ marks the inner boundary of the IP model, while the dashed white circle at $r=10.6$ indicates the outer boundary of the CO model. Note that due to slight differences in the output cadences of the two models, the results shown in the same panel may correspond to marginally different times.}
	\label{fig:CO_IP_cross}
\end{figure*}

\begin{figure*}
	\centering
  \includegraphics[width=\textwidth]{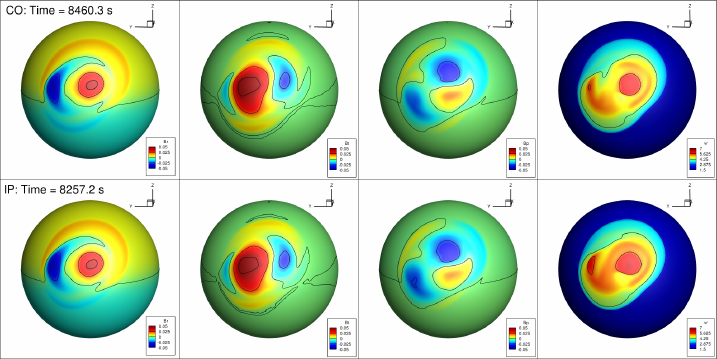}
	\caption{Comparison of MHD variables extracted from the CO and IP models on the sphere at $10R_\odot$. The quantities shown include the three magnetic field components ($B_r$, $B_\theta$, and $B_\phi$) and the plasma radial velocity $v_r$. Note that, again, there is slight difference in the times of the two models.  The close agreement between the two solutions confirms that the coupling method accurately transfers information from the corona to the inner heliosphere without introducing numerical artifacts at the interface.}
	\label{fig:compare10Rs}
\end{figure*}

\begin{figure*}
	\centering
  \includegraphics[width=\textwidth]{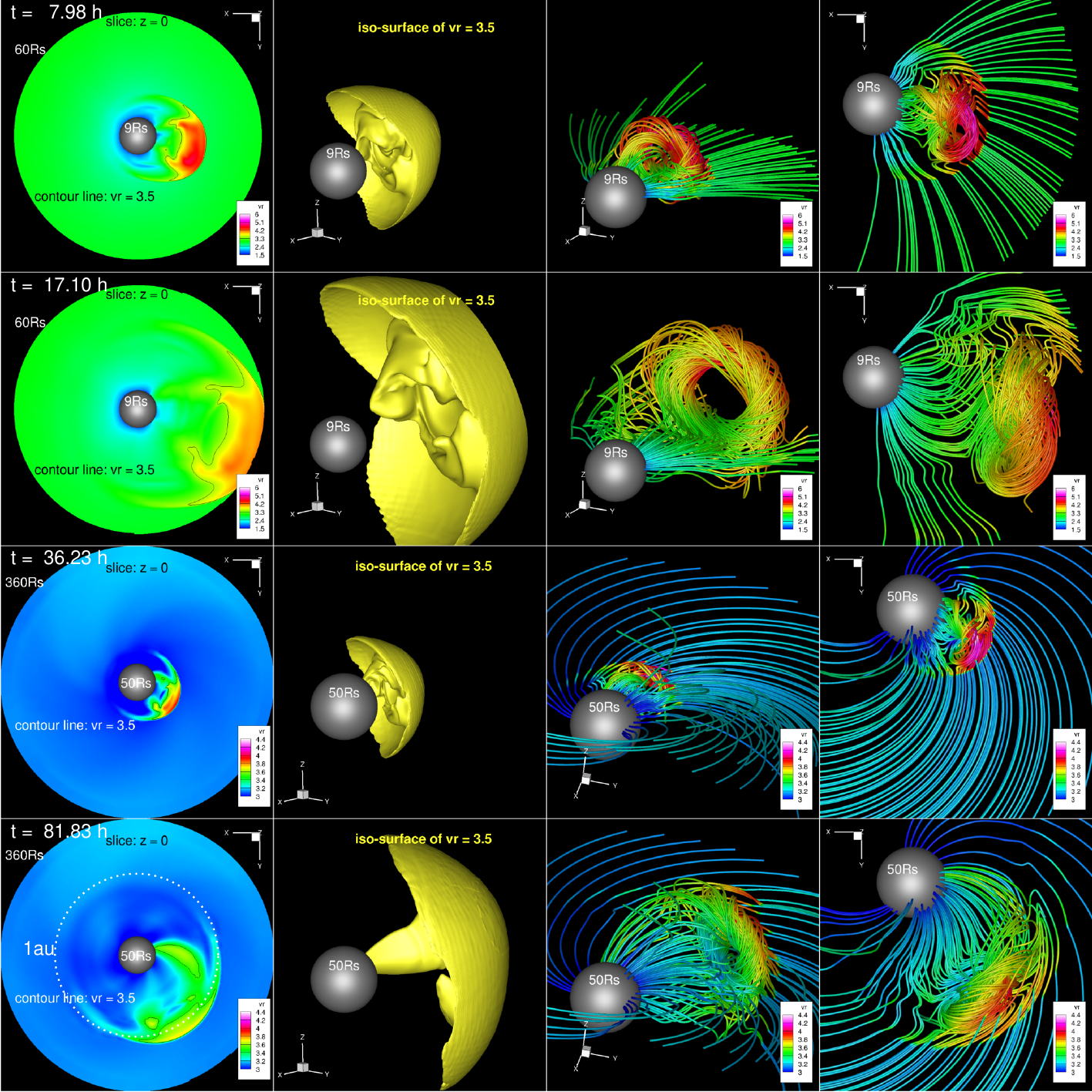}
	\caption{Propagation of the CME in the heliosphere. From left to right, the columns show the distribution of radial velocity $v_r$ in the equatorial plane, a three-dimensional iso-surface of $v_r$, and magnetic field lines viewed from two different angles: a three-dimensional perspective (third column) and a view from the north pole downward (fourth column). The magnetic field lines are color-coded by $v_r$. From top to bottom, the first two rows display results from the IP model covering the domain from $9R_{\odot}$ to $65R_{\odot}$, while the last two rows show results from the HE model extending from $50R_{\odot}$ to $360R_{\odot}$. In the left panel of the bottom row, the dashed circle indicates the location of 1~au. An animation of this evolution is available online, which shows the whole duration of the IP and HE simulations. The format of plot in panels of the animation is identical to those of this figure, and the top row is shown for IP and the bottom for HE.}
	\label{fig:cmepropa}
\end{figure*}

\begin{figure*}
	\centering
  \includegraphics[width=\textwidth]{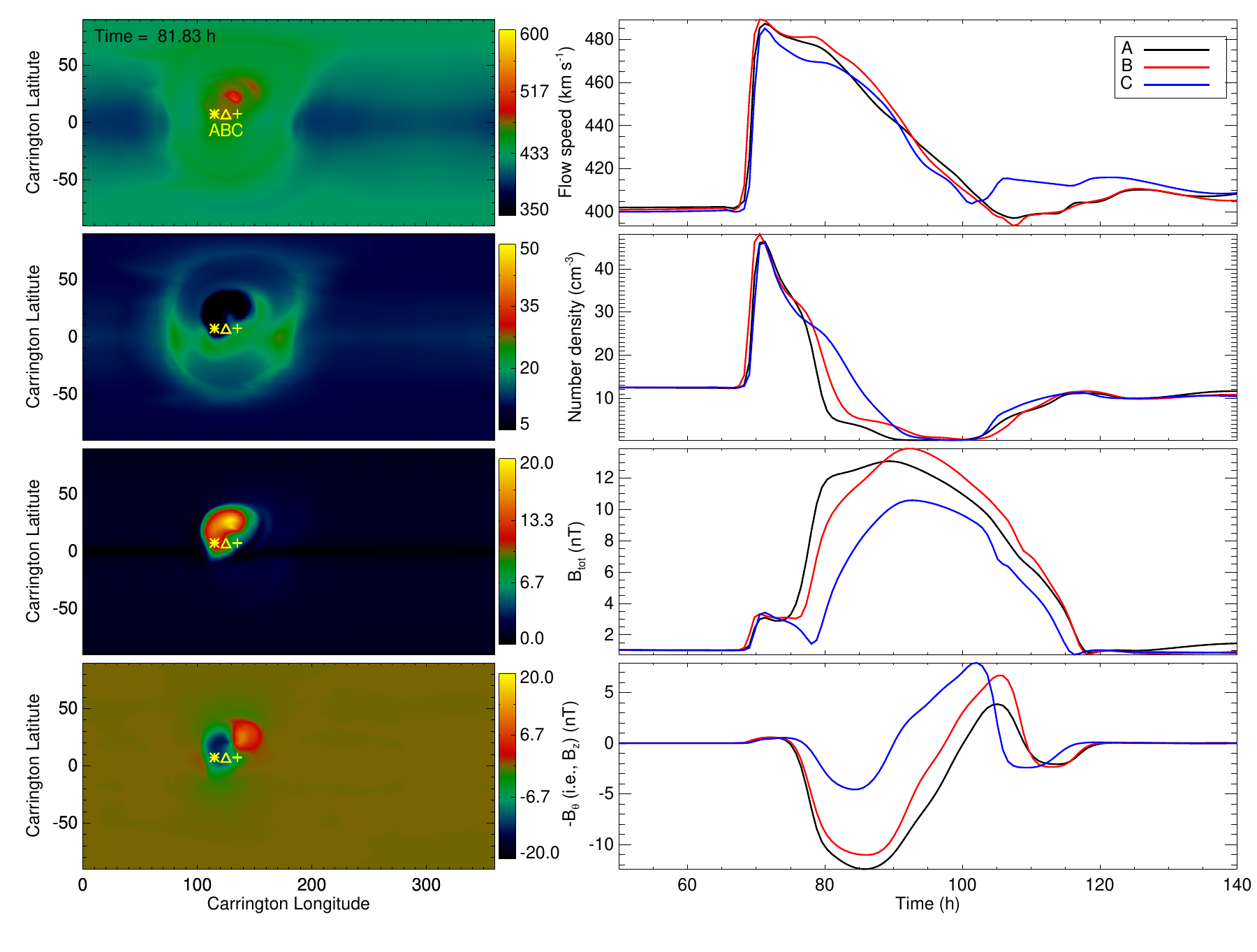}
	\caption{Synthetic in-situ observations at 1~au. Left panels: From top to bottom, the distributions of flow speed, plasma number density, total magnetic field strength, and the northward magnetic field component ($-B_{\theta}$) on the spherical surface at $r = 214.36 R_{\odot}$. Three sampling points mimicking the L1 point are labeled as A, B, and C, corresponding to different Carrington longitudes  ($160^\circ$, $170^\circ$, and $180^\circ$, respectively) of the solar disk center at $t = 0$ as seen on Earth. Right panels: Time profiles of the corresponding parameters extracted at these three points, showing the CME-driven disturbances as they propagate past Earth. An animation of this evolution is available online, which shows the whole duration of the simulation. The format of plot in panels of the animation is identical to those in this figure. Note that for time profiles of L1 points, only the northward magnetic field component is shown in the animation.}
	\label{fig:L1}
\end{figure*}

\subsection{CME propagation}
The CO model is run until the CME has completely exited its computational domain. During the later phase of the CO simulation, we gradually reduce the artificial light speed from its initial value of $c = 100$ down to $c = 30$, which is the same value used throughout the IP and HE models. This gradual reduction significantly accelerates the computation without affecting the physical evolution of the CME, as the eruption has already passed its main acceleration phase and the plasma speed has decreased considerably.

The CME takes approximately $80$ minutes to reach the outer boundary of the CO domain, after which it continues into the IP model. \Fig~\ref{fig:CO_IP_cross} shows the CME crossing the interface between the two models, displaying the normalized current density $J/B$, the radial velocity $v_r$, and the logarithm of density $\log \rho$ on a central meridional slice. The smooth transition across the interface, with no visible reflection or discontinuity, confirms the effectiveness of the coupling method.

To further validate the coupling, \Fig~\ref{fig:compare10Rs} compares the MHD variables extracted from both models on the spherical surface at $10R_{\odot}$. It is worth noting that this surface is located more than ten grid cells away from the driving surface extracted from CO (i.e., $9R_{\odot}$) and used in the IP model. The quantities are nearly identical, demonstrating that the information transfer from CO to IP preserves the solution accurately. These results confirm that the sequential coupling approach successfully maintains consistency between the models and allows the CME to propagate seamlessly from the corona into the inner heliosphere.

\Fig~\ref{fig:cmepropa} and the accompanying animation present the propagation of the CME in the heliosphere up to $360 R_{\odot}$. The CME front takes $70$ hours to reach 1~au, and shows a gradual deceleration during the travel. The whole structure shows an approximately self-similar expansion as it moves outward. The magnetic field of the CME mainbody demonstrates a donut-shaped structure (a torus), whose plane appears to become perpendicular to the solar equatorial plane. Note that the torus is not fully self-closed but with field lines still connecting to the Sun.

\Fig~\ref{fig:L1} and the accompanying animation present the distribution of key plasma and magnetic field parameters on the sphere at 1 au (more exactly the radius at L1 point with $r = 214.36R_{\odot}$), including the radial velocity, plasma density, total magnetic field strength, and the south-north magnetic field component ($-B_{\theta}$). To emulate in-situ observations at Earth, we extract time profiles at three points corresponding to different Carrington longitudes of the Earth at the start of the simulation ($t = 0$). These points are chosen such that the Carrington longitude of the solar disk center, as seen from Earth, is $160^\circ$ (point A), $170^\circ$ (point B), and $180^\circ$ (point C). In the latter case, the AR is located at the central meridian, since its center has a Carrington longitude of $180^\circ$. The latitude of all three points is fixed at $7.25^\circ$, corresponding to the typical heliographic latitude of Earth.

The time profiles show that upon reaching 1~au, the CME-driven shock front propagates at a speed of approximately $480$~km~s$^{-1}$, which is about $80$~km~s$^{-1}$ faster than the ambient solar wind. A sharp increase in plasma density is observed at the shock front, followed by a density depletion within the main body of the CME, corresponding to the magnetic flux rope (or magnetic cloud). Within this region, the magnetic field components exhibit a smooth rotation, characteristic of a flux rope structure. The total magnetic field strength reaches values between $10$ and $15$~nT. A southward $B_z$ component (which is approximate the $-B_\theta$) begins to appear approximately five hours after the arrival of the shock front and persists for over 20 hours in points A and B, with a peak value of around $13$~nT. The C point has a much shorter and weaker southward field. The five-hour interval between the shock front and the flux rope corresponds to the CME sheath region. The entire disturbance of the background solar wind caused by the CME lasts about 50 hours, spanning from $t = 70$~h to $t = 120$~h.

\begin{figure*}
	\centering
  \includegraphics[width=0.7\textwidth]{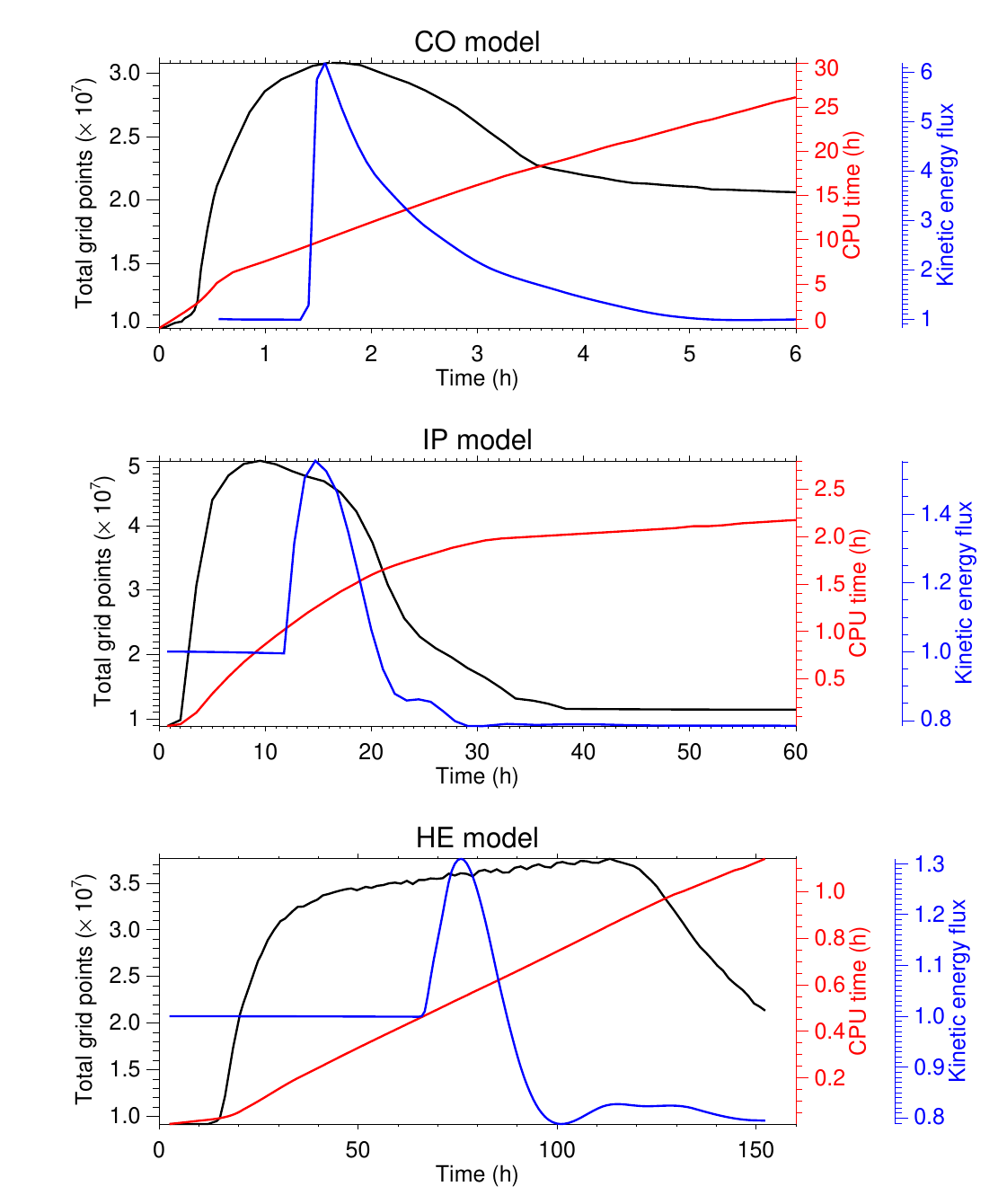}
	\caption{Computational cost and grid adaptation for the three coupled models. In each panel, the black curve shows the time evolution of the total number of grid points, the red curve represents the cumulative CPU time, and the blue curve indicates the kinetic energy flux at the outer boundary, which serves as a diagnostic for CME exit from the computational domain. All simulations were performed using 640 processors running at 2.6 GHz. From top to bottom, the panels correspond to the CO, IP, and HE models, respectively.}
	\label{fig:cputime}
\end{figure*}

\begin{figure*}
	\centering
  \includegraphics[width=0.9\textwidth]{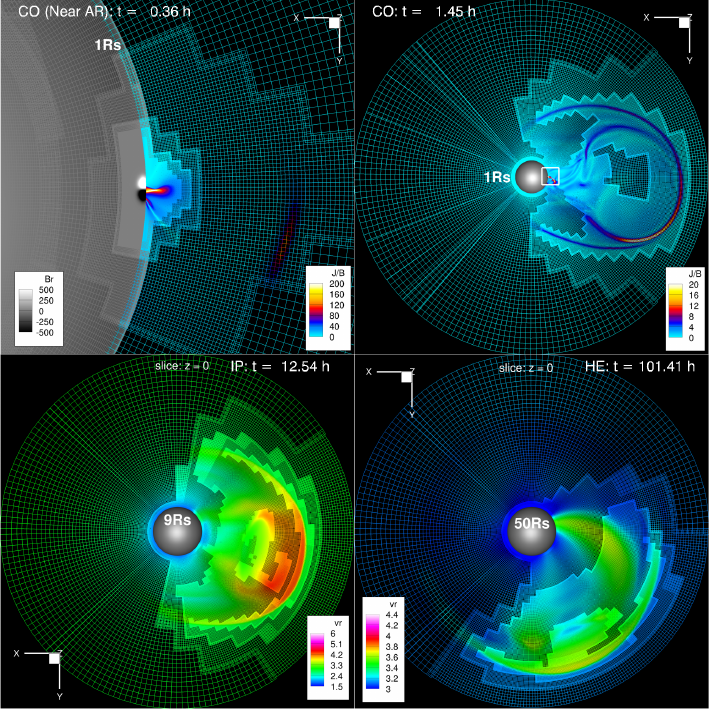}
	\caption{Adaptive grid structure of the simulation. The grid lines, formed by centers of all cells (including one layer of guard cells), are displayed on a central cross-section and color-coded according to the values indicated by the color bars in each panel. Four different times and fields of view are shown. The inner sphere in each panel is labeled with its corresponding radius. The top-left panel presents a zoomed-in view of the AR area (which is outlined by the white box in the top-right panel), and the solar surface is also shown with the grid lines. An animation covering the full simulation time is available online. The format of plot in panels of the animation is identical to those in this figure.}
	\label{fig:grid}
\end{figure*}

\subsection{Computation efficiency}
Computational efficiency is a critical factor for Sun-to-Earth CME simulations, particularly if the model is intended for operational space weather forecasting. \Fig~\ref{fig:cputime} illustrates the computational cost of the present simulation by showing the time evolution of the total number of grid points (black curves) and the cumulative CPU time (red curves) for each of the three coupled models. All computations were carried out using 640 processors running at $2.6$~GHz. Each model was run until the kinetic energy flux at its outer boundary became saturated, indicating that the CME and its associated disturbances had fully exited the domain. The adaptive refinement of the grid is illustrated in \Fig~\ref{fig:grid} (and the accompanying animation), which shows the grid structure on a central cross-section at different stages of the simulation. The refinement closely follows the CME front, the current sheet, and the developing flux rope, demonstrating that the AMR criteria successfully achieve the design goal of concentrating resolution where it is most needed while keeping the overall grid count manageable.

Correspondingly, the number of grid points evolves dynamically in response to the developing structures. In the CO model, for instance, the grid count increases sharply after eruption onset as the out-of-equilibrium magnetic field rapidly expands. Additional grid points are also triggered by the formation of the shock ahead of the CME and the elongation of the current sheet behind it. By the time the CME reaches the outer boundary of the CO domain, the total number of grid points peaks at approximately $3\times 10^7$, about three times the initial grid count. Throughout the entire simulation, the number of grid points across all models remains a few of $10^7$, as constrained by the AMR refinement criteria. The majority of the computational cost is incurred by the CO model, which requires about 26 CPU hours to complete the simulation up to $t = 6$~h, when the CME has fully left the domain. In contrast, the IP model uses only about 2 CPU hours to propagate the CME to $64R_\odot$, and the HE model requires roughly 1 CPU hour to track the CME to $360R_\odot$. In total, the entire Sun-to-Earth simulation of a typical CME from initiation to arrival at 1~au can be completed in approximately one day on a moderately sized parallel computing cluster.

This level of performance is particularly promising for forecasting applications. Given that CMEs typically take $3$-$4$ days to reach Earth, the model can provide predictions of key interplanetary parameters at 1~au with a lead time of about two days, making it a viable candidate for future operational space weather prediction systems.

\section{Conclusions}\label{sec:concl}

In this paper, we have presented a methodology for simulating the complete Sun-to-Earth evolution of a CME originating from a sheared magnetic arcade. The model couples three nested MHD simulations covering the corona (CO, $1$-$11R_{\odot}$), the interplanetary space (IP, $9$-$64R_{\odot}$), and the heliosphere (HE, $50$-$360R_{\odot}$), enabling seamless tracking of the CME from its birth near the solar surface to its arrival at 1~au. The main features of the model and the simulation results are summarized as follows.

The model incorporates three technical features that address common challenges in global CME simulations. 
(1) A block-structured AMR scheme is used to achieve multi-scale resolution, with the finest grid reaching $\sim 700$~km near the solar surface comparable to the resolution of SDO/HMI magnetograms. This allows the simulation to resolve fine-scale magnetic structures within ARs at scales consistent with observations. 
(2) A semi-relativistic Boris correction, combined with a relativistic mass-density factor, is implemented in the MHD equations. This modification limits the effective Alfv{\'e}n speed and permits the use of realistic active-region magnetic field strengths (up to several thousand Gauss) without requiring excessively small time steps. By appropriately adjusting the value of artificial light speed, the model can preserve the physical acceleration and dynamics of eruptions while maintaining computational feasibility. (3) A modular approach is developed to couple the three models through prescribed inner boundary conditions. The CO model provides time-dependent MHD variables at $9R_{\odot}$ to drive the IP model, which in turn drives the HE model at $50R_{\odot}$. The coupling method ensures consistency of plasma parameters and magnetic field evolution across the interfaces, and validation tests show no numerical reflection or discontinuity at the boundaries.

Using this modeling framework, we have simulated the entire life cycle of a CME initiated from a sheared arcade. The pre-eruption phase is characterized by quasi-static energy buildup driven by photospheric shearing motions, leading to the formation of a current sheet and subsequent magnetic reconnection that triggers the eruption. The CME rapidly accelerates, reaching speeds over $3000$~km~s$^{-1}$ in the low corona, and then propagates through the heliosphere with gradual deceleration. The simulated CME exhibits a classic three-part structure in synthetic coronagraph images, with a bright front, dark cavity, and bright core. In the heliosphere, the CME main body evolves into a donut-shaped magnetic flux rope whose orientation rotates during propagation. Synthetic in-situ observations at 1~au show that the CME arrival time is approximately $70$ hours after eruption, with a shock speed of $\sim 480$~km~s$^{-1}$, density compression, and a prolonged southward $B_z$ component up to $13$~nT.

The computational efficiency of the model has also been assessed. The entire Sun-to-Earth simulation, from eruption onset to CME passage beyond $360R_{\odot}$, requires approximately one day on a moderately sized parallel cluster with 640 processors. This level of performance is of practical interest for space weather forecasting applications, as it provides a lead time of about two days relative to the typical $3$-$4$ day CME transit time to Earth.

Further analysis of the simulation results will be presented in a companion paper, where we will examine in detail the three-dimensional structure of the CME magnetic field, particularly the formation mechanism of the flux rope, and its evolution and propagation processes. In future works, on the one hand, we plan to apply this modeling framework to study more complex scenarios, such as the generation of homologous CMEs driven by continuous shearing motions~\citep{bianHomologousCoronalMass2022} and the interaction of the resulting multiple CMEs. On the other hand, we plan to incorporate more realistic thermodynamic processes into the model to reproduce the fast and slow solar wind streams, thereby obtaining a more accurate background solar wind. Furthermore, we aim to drive the AR evolution using observed vector magnetograms, enabling a data-driven MHD simulation of the initiation, evolution, and propagation of CMEs~\citep{jiangDatadrivenModelingSolar2022}. The ultimate goal is to develop a data-driven MHD modeling framework capable of predicting interplanetary space weather parameters and assessing the geoeffectiveness of Earth-directed CMEs.

\begin{acknowledgments}
  This work is jointly supported by National Key R\&D Program of China under No. 2024YFA1612001, National Natural Science Foundation of China (NSFC 12573058), Shenzhen Science and Technology Program (grant no. RCJC20210609104422048), Guangdong Basic and Applied Basic Research Foundation (2023B1515040021), and the Specialized Research Fund for State Key Laboratory of Solar Activity and Space Weather.
\end{acknowledgments}


\end{document}